\documentclass[aps,prl,letterpaper,twocolumn,nofootinbib,floatfix,showpacs]{revtex4}

\usepackage{graphics}
\usepackage{amsmath}
\usepackage{epsfig}

\begin{document}



\title{Electronic signature of DNA nucleotides via transverse transport}

\renewcommand{\thefootnote}{\fnsymbol{footnote}}
\author{Michael Zwolak}
\affiliation{Physics Department, California Institute of Technology, Pasadena, California 91125}
\author{Massimiliano Di Ventra\cite{MD}}
\affiliation{Department of Physics, University of California, San Diego, La Jolla, CA 92093-0319}
\date{\today}
\begin{abstract}
%
We report theoretical studies of charge transport in single-stranded DNA in the
 direction perpendicular to the backbone axis. We find that, if the electrodes
which sandwich the DNA have the appropriate spatial width, each nucleotide
carries a unique signature due to the different electronic and chemical
structure of the four bases. This signature is independent of the 
nearest-neighbor nucleotides. Furthermore, except for the nucleotides with
Guanine and Cytosine bases, we find that the difference in conductance of the
nucleotides is large for most orientations of the bases with respect to the
electrodes. By exploiting these differences it may be possible to sequence
single-stranded DNA by scanning its length with conducting probes.
\end{abstract}

\maketitle

There has recently been an increased interest in charge transport in 
DNA, due to both its relevance in physiological reactions and to its potential 
use in molecular electronics.~\cite{ourreview,barton1,revmod} Previous studies have looked into the effect of the base sequence 
and structural distortions on charge transport, and the interplay between different transport 
mechanisms.~\cite{ourreview,revmod,giese1,lewis,schuster,li,berlin,hjort,zhang,roche,grozema} 
However, much of the research so far has focused 
on how charge flows {\it along} the DNA axis. Very few experimental studies have looked into the transport properties of DNA in 
the {\it transverse} direction. For instance, scanning probe techniques have been 
used to image isolated DNA bases or a few bases in single-stranded DNA~\cite{tao,tanaka,hamai,tanaka1,wang} 
but no systematic study is available on the relative ability of each base to carry electrical current. 

\begin{figure}
\begin{center}
\includegraphics*[width=7.5cm]{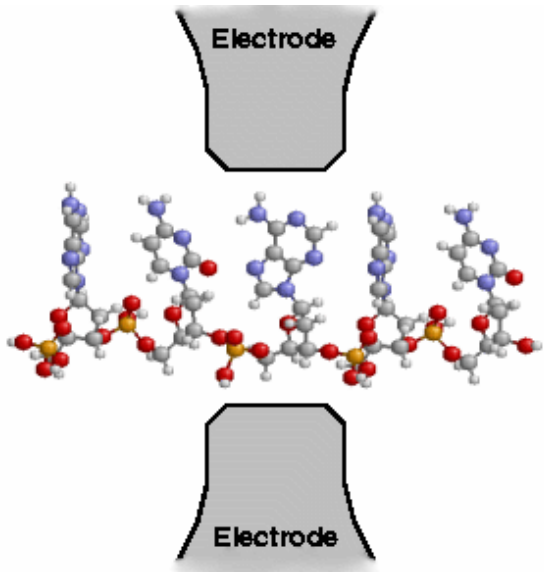}
\caption{Schematic of a polynucleotide between two electrodes. The electrodes only couple 
to a single nucleotide at a time. }
\label{schematic}
\end{center}
\end{figure}

In this paper, we report on charge transport in single-stranded DNA in the direction {\em perpendicular} to 
the backbone axis. Our goal is to determine the relative current carried by each nucleotide in the presence of its neighbors. We envision a possible experimental setup as schematically shown in Fig.~\ref{schematic} where two probes of 
nanoscale dimensions sandwich the DNA in such a way that one or a few bases are spatially confined within the lateral 
dimension of the probes. Experimentally, this could be realized by a polynucleotide 
molecule between two electrodes embedded in a nanopore~\cite{pore1,pore2} or by adsorbing the polynucleotide 
onto a metal surface and using a scanning tunneling microscope (STM) tip to measure the current through a single nucleotide.~\cite{tanaka,hamai} 
We therefore seek to identify specific electronic signatures of each nucleotide that would allow us to 
create a {\it DNA electronic map}. In particular, we investigate the role 
the electronic and chemical structures of the bases play in the charge transport, as well as the role of 
interactions between each base and its nearest neighbors.
We indeed find that the current through each individual nucleotide provides a unique electrical 
signature of each base. The ratio of the current for different nucleotides gives a measure of the difference 
of their signatures. These electrical signatures change with orientation of the nucleotides, but are only affected 
slightly by neighboring nucleotides so long as the electrode lateral dimension is comparable with 
the average ``size'' of the bases (i.e., a dimension on the order of 1.0 nm, comparable to the nucleotide spacing 
of 0.7-0.9 nm~\cite{hamai}). In most cases, the current ratios are quite large when comparing between 
nucleotides of random orientations, except for the 
ratio of the current through 2$^{\prime}$-deoxyguanosine 5$^{\prime}$-monophosphate and the current through 
2$^{\prime}$-deoxycytidine 5$^{\prime}$-monophosphate (i.e., nucleotides with
guanine and cytosine bases). Such findings may help create a practical sequencing tool based on the electronic read-out of DNA.  

We consider a single strand of DNA, e.g., a polynucleotide, between two electrodes kept at a given bias,  
(see schematic in Figure~\ref{schematic}). The electrodes are represented by 3 layers of about 30 atoms arranged in 
the [111] surface geometry, which gives an electrode surface spanning about 100~\AA$^2$. The electrode-electrode spacing 
is fixed at 15~\AA. Both the area of the electrode surface and the electrode-electrode 
distance are such that the largest of the nucleotides (i.e., the one with the A base) can be 
accommodated between the electrodes. Since we are interested only in the {\it ratio} 
between currents of different nucleotides (and we consider only linear response), we 
employ a simple tight-binding model for the electronic structure of the system with $s, p_x, p_y,$ and $p_z$ orbitals
for each carbon, nitrogen, oxygen, and phosphorus atom; and $s$ orbitals for hydrogen and gold. We then use a Green's function 
approach in order to calculate the current through the whole system (electrodes plus 
DNA).~\cite{datta,nanotube} 
The Green's function $\mathcal{G}_{DNA}$ of the combined electrode-DNA-electrode 
system is written as
\begin{equation}
\mathcal{G}_{DNA}(E)=[E\mathcal{S}_{DNA}-\mathcal{H}_{DNA}-\Sigma_t-\Sigma_b]^{-1}
\end{equation}
where $\mathcal{S}_{DNA}$ and $\mathcal{H}_{DNA}$ are the overlap and the Hamiltonian 
matrices, respectively. $\Sigma_{t(b)}$ are the self-energy terms that describe the effect 
of the electronic structure of the leads. 

The transmission coefficient $T(E)$ is given by~\cite{datta}
\begin{equation}
T(E)=T_{tb}=\mathrm{Tr}[\Gamma_t\mathcal{G}_{DNA}\Gamma_b\mathcal{G}_{DNA}^\dagger]
\end{equation}
where $\Gamma_{t(b)}=i(\Sigma_{t(b)}-\Sigma_{t(b)}^\dagger)$. We calculate the current from 
\begin{equation}
I = \frac{2 e}{h} \int_{-\infty}^{\infty} dE T(E) [f_t(E) -f_b(E)], 
\end{equation}
where $f_{t(b)}=\{\exp[(E - \mu_{t(b)})/k_BT] + 1\}^{-1}$ is the Fermi function, 
and $\mu_{t(b)}$ is the chemical potential of the top (bottom) electrode (see Fig~\ref{schematic}). We 
assume the voltage drops equally in the space between the top electrode and the 
molecule, and the molecule and bottom electrode. The current is calculated  
at room temperature.~\cite{prec}

\begin{figure}
\begin{center}
\includegraphics*[width=7.5cm]{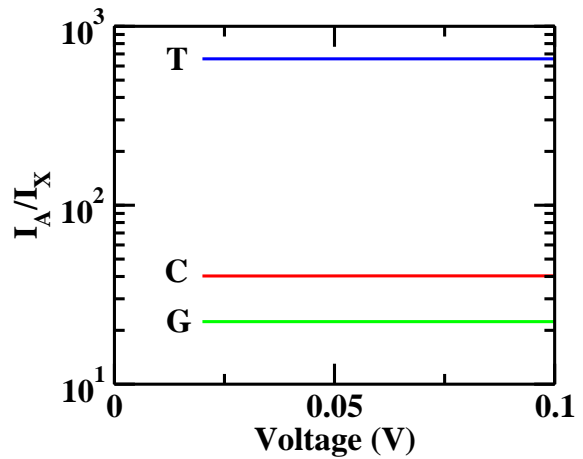}
\caption{Current ratios of A to the other nucleotides $X$ ($X=$ G, C, or T) up 
to a bias of $0.1 V$. }
\label{factors}
\end{center}
\end{figure}

We first examine the current ratios for the four nucleotides (which we 
label by A, G, C, or T corresponding to their base) without nearest-neighbor 
nucleotides attached. The nucleotides are therefore ``passivated'', i.e., the oxygen 
atoms which are normally charged in solution are instead bonded to a hydrogen atom.~\cite{ourreview} 
For each nucleotide, we relaxed the geometry using total-energy Hartree-Fock 
calculations.~\cite{prec1} We consider first the nucleotides positioned so that the base is in a plane 
perpendicular to the electrode surface (see Fig.~\ref{schematic}). Figure~\ref{factors} shows the ratio between the current of 
A and the current of the other nucleotides, $I_{A}/I_{X}$ where $X=$ G, C, or T. The ratios are about 20, 40, and 660 for 
G, C, and T, respectively. Strong contrast has indeed been observed in an STM experiment between A 
and T nucleotides.~\cite{tanaka}
Figure 2 thus indicates that the isolated 
nucleotides can be distinguished relatively easily from one another using transverse currents. 

The differences in the electrical current ratios are due to the relative position of the Fermi level with respect to the 
highest-occupied molecular orbital (HOMO) and the lowest-unoccupied molecular orbital (LUMO) as well as the value of the density of 
states (DOS)~\cite{miolang} of the base at the Fermi level. By projecting onto the different atomic orbitals, we find that the states 
of the bases make up for most of the HOMO and LUMO levels while the DOS at the Fermi level is mainly determined by the 
backbone states due to the large coupling of the backbone with the bottom electrode. However, the relative current is 
closely related to the value of the DOS at the Fermi level due to the bases, which gives the ordering 
A $>$ G $\sim$ C $>$ T. How well those base states couple to {\it both} electrodes determines the overall magnitude of the relative currents.

\begin{figure}
\begin{center}
\includegraphics*[width=7.5cm]{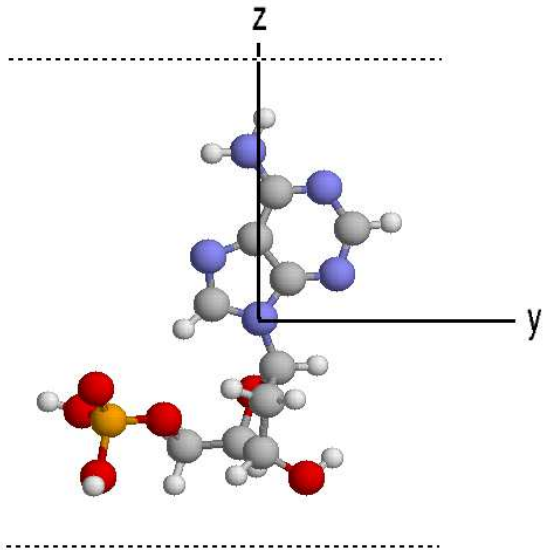}
\caption{Coordinate system which we use to define rotations and translations of the nucleotides 
with respect to the probe surfaces. The 
x-axis is pointing out of the paper. The dashed lines represent the position of the electrode surfaces.}
\label{coor}
\end{center}
\end{figure}

\begin{figure}
\begin{center}
\includegraphics*[width=7.5cm]{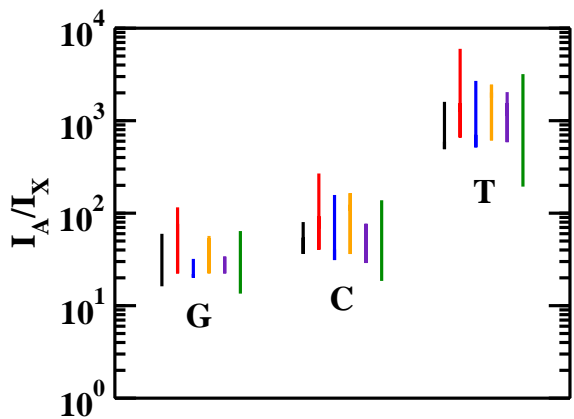}
\caption{Current ratios of A to the other nucleotides $X$ ($X=$ G, C, or T) at
a bias of $0.1 V$. $I_A$ is for A at its original configuration, while the 
other nucleotides' orientations are varied. The 6 lines for each nucleotide, from left 
to right, correspond to rotation about the x-, y-, and z-axes
and translation in the x-, y-, and z-directions.}
\label{var2}
\end{center}
\end{figure}

Having shown that each nucleotide carries a different electrical signature for a particular geometry, we can now examine how this result changes with orientation of the base with respect to the probe surfaces and location of the backbone atoms relative to the surface. 
We define our coordinate system as shown in Figure~\ref{coor}. 
The origin is the nitrogen which attaches each base to the 
phosphate-sugar backbone. Both rotation and translation involve a rigid movement of the whole nucleotide (base 
plus backbone). The x-axis runs parallel to the electrode surfaces and 
perpendicular to the plane of the base. The y-axis runs parallel to the electrode surfaces 
and in the plane of the base.  The z-axis runs perpendicular to the electrode 
surface. We consider rotations and translations of the DNA nucleotides about each of the axes 
independently. The six variations we consider are the following: (1) rotation about the x-axis from -10 to
+10 degrees, (2) rotation about the y-axis from -15 to 15 degrees, (3) rotation about the
z-axis from -30 to +30 degrees, (4) translations along the x-axis by -1.5 to 1.5~\AA,
(5) translations along the y-axis by -1.5 to 1.5~\AA, and (6) translation along the z-axis by
-0.5 to 0.5~\AA \ about the original nucleotide position.~\cite{prec2} 

Figure~\ref{var2} shows relative changes in current with nucleotide rotation and 
translation as defined above. For each nucleotide, the lines from left to right correspond to rotation about the x-, y-, and z-axes and 
to translation along the x-, y-, and z-axes, respectively. 
The variations (2) and (6) (i.e., rotation about the y-axis and translation along the z-axis, respectively) cause the largest 
changes in relative current. This is easy to understand since both movements change the 
distance between the uppermost atoms on the nucleotides and the electrode surface, 
which gives a large change in coupling. 
This is analogous to the double-stranded DNA case: variation of the angle and distance 
between bases can significantly 
change the conductivity because of a reduction of the inter-base coupling.~\cite{grozema} From Fig.~\ref{var2}, 
it is clear that, even taking into account the above variations, there is a large 
distinction between the nucleotides A and T and the pair \{G,C\}. The 
latter two have a similar electrical signature for a wide range of rotations and translations and cannot therefore 
be easily distinguished. Shot noise, however, being more sensitive to changes in electronic structure than the 
current,~\cite{chennoise} could be used to distinguish between the latter two bases. We leave this study for 
future investigations. 

We conclude by looking at the effect of nearest-neighbor nucleotides on the electrical signature of each 
individual nucleotide. 
It is known from ground-state calculations~\cite{ladik,otto,clem} that in 
isolated stacks of nucleotides the states on each base are highly localized on that base. 
We therefore expect small corrections due to nearest-neighbor base interactions. We have checked this point by looking at the following representative sequences of three nucleotides of the type A$X$A, A$X$T, C$X$A, G$X$G, G$X$T, and T$X$T, where 
$X=$ G, C, or T is the nucleotide that is sandwiched between the two electrodes.~\cite{prec3} 
The base of the central nucleotide is also assumed perpendicular to the electrode surface.

\begin{figure}
\begin{center}
\includegraphics*[width=7.5cm]{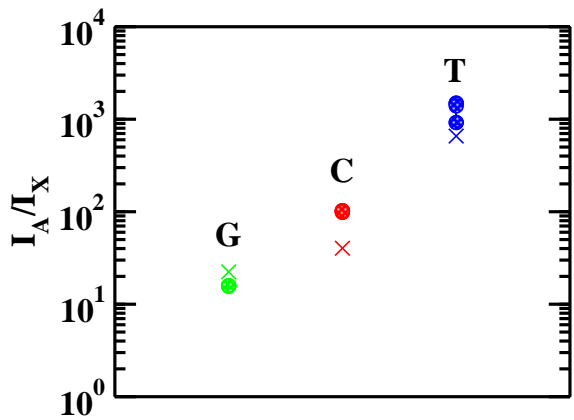}
\caption{Current ratios for strands of nucleotides with central nucleotide $X$ ($X=$ G, C, or T) at
a bias of $0.1 V$. $I_A$ is for isolated A at its original configuration.
 The crosses are for the individual nucleotides and the circles are for sequences 
of three nucleotides (see text for details).} 
\label{strandIV}
\end{center}
\end{figure}

The relative change in current for these cases is shown in 
Figure~\ref{strandIV} at 0.1V. As expected, there is little change in current compared to the 
individual nucleotides so long as the spatial width of 
the electrodes is only large enough to accommodate just one base (see Fig.~\ref{schematic}). 
Most of the change is due to the difference in backbone geometry of the strand 
compared to the isolated nucleotide. Although, for sequences with T as the central nucleotide, the current changes 
due to nearest neighbors: T carries the smallest current and therefore is the most sensitive to its nearest 
neighbors interacting with the electrodes. We therefore conclude that the electrode width needs to be about 1 nm 
if the nearest neighbor bases have uncorrelated orientations. When this is the case, there are variations due to 
single nucleotide orientation, as shown in Figure~\ref{var2}, but there is not significant variations due nearest neighbor interaction.  

In conclusion, we have studied charge transport through nucleotides 
in the direction transverse to the backbone axis. 
We find that each nucleotide has a unique electrical signature determined by the electronic and chemical 
properties of the 
bases irrespective of the nearest-neighbor nucleotides. Except for 
the nucleotides with Guanine and Cytosine bases, these signatures are quite robust with variation in nucleotide orientation and location with respect to the electrodes. 
By exploiting these differences, it may be 
possible to sequence DNA by scanning its length with a conducting probe. Clearly, for such application 
many other factors (like solvent and ionic effects) need to be studied in detail and 
additional theoretical work is necessary. Also, the full noise spectrum may provide extra diagnostic 
capability to differentiate between bases. 

One of us (MD) is indebted to Len Feldman and Mike Ramsey for their suggestion to work on this topic and for useful 
discussions. We acknowledge partial support from the NSF 
Grant DMR-01-33075 and the National Human Genome Research Institute. One of us (MZ) also 
acknowledges support from an NSF Graduate Fellowship.


\begin{thebibliography}{99}
\bibitem{prec1} The relaxed structure of each portion of the backbone which is attached to a 
base is not a local property of the nucleotide, but is instead a property of the nearby 
sequence and the environment. We therefore first completely relaxed the 2$^{\prime}$-deoxyadenosine 5$^{\prime}$-monophosphate, from 
which we took the relaxed phosphate-sugar structure. We then employed this structure for all other nucleotides. We finally relaxed the 
latter nucleotides keeping the phosphate-sugar 
atoms fixed. This also allows each nucleotide to be 
positioned between the electrodes in a similar way, which ensures that differences in electrical 
current are due to the bases only.
\bibitem{prec2} The choice of the ranges for variations (1) and (6) is dictated by the size of the electrodes, 
i.e., in an actual experiment and for 
a polynucleotide, the 
x-axis rotation and z-axis translation are bound by the distance between probes. The ranges for variations (3), 
(4) and (5) are limited by the geometry of the [111] surface. The rotation around the y-axis is, 
on the other hand, arbitrary unless the spatial direction along the x-axis is also limited, as in, e.g., 
a nanopore.
\bibitem{prec3} The coordinates of these sequences have been relaxed by total-energy Hartree-Fock 
calculations in a similar way to the isolated nucleotides (see Ref.~\onlinecite{prec1}).
\end{thebibliography}
\end{document}